\documentclass[11pt,a4]{article}
\usepackage{pstricks}

\usepackage{color}

\def\half{\frac{1}{2}}

\newfont{\bbbold}{msbm10 scaled \magstep1}

\def\cL{{\cal L}}

\def\cN{{\cal N}}
\def\cO{{\cal O}}

\newfont{\goth}{eufm10 scaled \magstep1}

\def\a{\alpha}
\def\b{\beta}\def\bdt{\dot \beta}
\def\c{\gamma}\def\C{\Gamma}\def\cdt{\dot\gamma}
\def\d{\delta}
\def\ve{\varepsilon}

\def\h{\eta}
\def\i{\iota}

\def\l{\lambda}\def\L{\Lambda}
\def\m{\mu}

\def\th{\theta}

\def\be{\begin{equation}}\def\ee{\end{equation}}
\def\bea{\begin{eqnarray}}\def\eea{\end{eqnarray}}
\def\barr{\begin{array}}\def\earr{\end{array}}

\def\o{\omega}
\def\del{\partial}

\def\uo{\underline\omega}
\def\una{\underline a}\def\unA{\underline A}

\def\unF{\underline F}

\def\unK{\underline K}

\def\unP{\underline P}

\def\unH{\underline{H}}
\def\unF{\underline{F}}
\def\unK{\underline{K}}
\def\unW{\underline{W}}

\def\xz{\times}

\def\nab{\nabla}

\let\la=\label

\def\nn{\nonumber}
\def\bd{\begin{document}}
\def\ed{\end{document}}
\def\ba{\begin{array}}
\def\ea{\end{array}}
\def\bea{\begin{eqnarray}}
\def\eea{\end{eqnarray}}
\def\ft#1#2{\tfrac{#1}{#2}}
\def\fft#1#2{\frac{#1}{#2}}
\def\sst#1{{\scriptscriptstyle #1}}
\def\oneone{\rlap 1\mkern4mu{\rm l}}

\newcommand{\eq}[1]{(\ref{#1})}
\newcommand{\w}[1]{\\[0.#1cm]}
\def\eqs#1#2{(\ref{#1}-\ref{#2})}
\def\det{{\rm det\,}}
\def\tr{{\rm tr}}

\newcommand{\hoch}[1]{$\, ^{#1}$}
\newcommand{\imperial}{\it\small Theoretical Physics Group, Imperial College London\\ Prince Consort Road, London SW7 2AZ, UK}
\newcommand{\kings}
{\it\small Department of Mathematics, King's College, University of London\\ Strand, London WC2R 2LS, UK}
\newcommand{\uu}
{\it\small Theoretical Physics, Department of Physics and Astronomy, Uppsala University, \\ Box 516, SE-75120, Uppsala, Sweden.}
\newcommand{\tamphys}{\it\small George and Cynthia Mitchell Institute for Fundamental Physics and Astronomy,\\
Texas A\&M University, College Station,
TX 77843, USA}
\newcommand{\golm}
{\it\small AEI, Max Planck Institut f\"ur Gravitationsphysik\\ Am M\"{u}hlenberg 1, D-14476 Potsdam, Germany}
\newcommand{\cpht}
{\it\small Centre de Physique Th\'eorique (UMR 7644), Ecole Polytechnique\\ 91128 Palaiseau Cedex, France}

\makeatletter
\@addtoreset{equation}{section} \makeatother

\newcommand{\sa}{/ \hspace{-1.2ex}}
\newcommand{\saa}{/ \hspace{-1.4ex}}
\newcommand{\saaa}{\, / \hspace{-1.6ex}}
\newcommand{\Scal}[1]{\Bigl ({#1} \Bigr )}
\newcommand{\scal}[1]{\bigl ({#1} \bigr )}

\newcommand{\CR}{\nonumber \\*}

\newcommand{\trace}{\hbox {tr}~}
\newcommand{\traceS}{\hbox {tr}_{\scriptscriptstyle \mathfrak{S}}~}

\DeclareMathAlphabet{\mathpzc}{OT1}{pzc}{m}{it}
\def\BRST{\,\mathpzc{s}\,}
\def\aBRST{{\scriptstyle (\mathpzc{s})}}
\def\q{{{\scriptscriptstyle (Q)}}}
\def\qs{{\scriptscriptstyle (Q\mathpzc{s})}}
\def\Qsla{{\mathcal{S}_{\q}}}
\def\Slav{{\mathcal{S}_\aBRST}}
\def\epsilonb{{\overline{\epsilon}}}
\def\bulletup{{\scriptstyle \bullet}}

\newcommand{\gra}[2]{{\scriptscriptstyle (#1 , #2 )}}
\newcommand{\ord}[1]{{\scriptscriptstyle (#1)}}

\def\cL{{\cal L}}
\def\cN{\mathcal{N}}
\def\cO{\mathcal{O}}

\def\ie{{\it i.e.}\ }
\def\eg{{\it e.g.}\ }

\newcommand{\sfrac}[2]{{\scriptstyle \frac{#1}{#2}}}
\newcommand{\stfrac}[2]{{\scriptscriptstyle \frac{#1}{#2}}}

 \def\balpha{{\overline{\alpha}}}
 \def\bbeta{{\overline{\beta}}}
 \def\bgamma{{\overline{\gamma}}}
 \def\bdelta{{\overline{\delta}}}
 \def\bepsilon{{\overline{\epsilon}}}
 \def\bvarepsilon{{\overline{\varepsilon}}}
 \def\bzeta{{\overline{\zeta}}}
 \def\bareta{{\overline{\eta}}}
 \def\btheta{{\overline{\theta}}}
 \def\bvartheta{{\overline{\vartheta}}}
 \def\biota{{\overline{\iota}}}
 \def\bkappa{{\overline{\kappa}}}
 \def\blambda{{\overline{\lambda}}}
 \def\bmu{{\overline{\mu}}}
 \def\bnu{{\overline{\nu}}}
 \def\bxi{{\overline{\xi}}}
 \def\bpi{{\overline{\pi}}}
 \def\brho{{\overline{\rho}}}
 \def\bvarrho{{\overline{\varrho}}}
 \def\bsigma{{\overline{\sigma}}}
 \def\bvarsigma{{\overline{\varsigma}}}
 \def\btau{{\overline{\tau}}}
 \def\bphi{{\overline{\phi}}}
 \def\bvarphi{{\overline{\varphi}}}
 \def\bchi{{\overline{\chi}}}
 \def\bpsi{{\overline{\psi}}}
 \def\bomega{{\overline{\omega}}}

\def\thalf{{\textrm{\tiny\textonehalf}}}
\def\tquarter{{\textrm{\tiny\textonequarter}}}
\def\Ko{{\scriptscriptstyle K}}
\def\tKo{\scriptscriptstyle k }
\def\corr{$\clubsuit$}
\renewcommand\theequation{\thesection.\arabic{equation}}

\newcommand{\auth}{\large G.\ Bossard\hoch{1}\hoch{2}, P.S.\ Howe\hoch{3}, U. Lindstr\"om\hoch{4}, K.S.\ Stelle\hoch{5} and L. Wulff\hoch{6}}

\begin{document}

\renewcommand{\thefootnote}{\fnsymbol{footnote}}

\null
\begin{flushright}
{\small AEI-2010-142}\\
{\small KCL-MTH-10-09}\\
{\small UUITP-27/10}\\
{\small Imperial/TP/10/KSS/02}\\
{\small MIFPA-10-53}\\
\vskip 1.5 cm
\end{flushright}

\begin{center}
{\Large{\bf Integral invariants in maximally supersymmetric Yang-Mills theories}}
\vspace{.75cm}

\auth

\vspace{.5cm}

\begin{itemize}
\item [$^1$]\golm\item[$^2$]\cpht\item[$^3$] \kings \item [$^4$] \uu \item [$^5$] \imperial\item[$^6$] \tamphys
\end{itemize}

\vspace{1cm}

{\bf Abstract}
\end{center}
\vskip .5cm
Integral invariants in maximally supersymmetric Yang-Mills theories are discussed in spacetime dimensions $4\leq D\leq 10$ for $SU(k)$ gauge groups. It is shown that, in addition to the action, there are three special invariants in all dimensions. Two of these, the single- and double-trace $F^4$ invariants, are of Chern-Simons type in $D=9,10$ and BPS type in $D\leq 8$, while the third, the double-trace of two derivatives acting on $F^4$, can be expressed in terms of a gauge-invariant super-$D$-form in all dimensions.  We show that the super-ten-forms for $D=10$ $F^4$ invariants have interesting cohomological properties and we also discuss some features of other invariants, including the single-trace $d^2 F^4$, which has a special form in $D=10$. The implications of these results for ultra-violet divergences are discussed in the framework of algebraic renormalisation.

\line(1,0){450}

{\sl \small email: bossard@cpht.polytechnique.fr, paul.howe@kcl.ac.uk,  ulf.lindstrom@fysast.uu.se, k.stelle@imperial.ac.uk, \linebreak linus@physics.tamu.edu}

\renewcommand{\thefootnote}{\arabic{footnote}}
\setcounter{footnote}{0}

\pagebreak\tableofcontents\setcounter{page}{1}


\section{Introduction}


Integral invariants in maximally supersymmetric theories, supergravity (MSG) or super Yang-Mills (MSYM), can be viewed as possible higher-order corrections to string or brane effective actions or as potential field-theoretic counterterms. Since no off-shell supersymmetric actions are known for maximal theories, these invariants have to be constructed in a perturbative fashion starting with expressions that are invariant under on-shell supersymmetry. Broadly speaking, there are  two main categories of on-shell invariants, those that can be expressed as integrals over the full superspaces (with 16 or 32 odd coordinates for MSYM and MSG respectively) of gauge-invariant integrands, and those that cannot. The latter can usually be expressed as integrals over some subsuperspace, that is as superactions \cite{Howe:1981xy}, or as generalised chiral (harmonic superspace) integrals \cite{Galperin:1984av}. We shall refer to the former as long and the latter, of which there are very few, as short, since the multiplets of which the invariants are the top components have these properties. We can regard any independent MSYM invariant as a possible deformation of the usual MSYM action. The presence of such a deformation will alter the supersymmetry transformations and induce higher-order terms as a consequence. In the case of full superspace integrals, there is no problem in extending such an invariant to all orders in an expansion parameter, such as $\a'$, but this is not so obvious for short invariants.

In the current paper we investigate the question of invariants for MSYM in spacetime dimensions $4\leq D\leq 10$, although we do not study their non-linear higher-order extensions. There is a full  classification for $D=4, N=4$ \cite{Drummond:2003ex}. This  is a special case because the starting theory in $D=4$ is superconformal, a fact that was exploited in \cite{Drummond:2003ex} in the construction of the invariants. The short invariants, of which there are just three, give rise to spacetime integrals that are the supersymmetric completions of terms of the form $\tr F^4, \tr^2 F^4$ and $d^2\tr^2 F^4$, where $\tr^2$ indicates a double-trace. The pure $F^4$ terms are one-half BPS, that is they correspond to short multiplets that are independent  of one-half of the odd coordinates when the latter are appropriately chosen, while the $d^2\tr^2 F^4$ invariant is one-quarter BPS, i.e. it is independent of one-quarter of the odd coordinates. However, it is sometimes useful to think of the latter as ``pseudo'' one-half BPS, that is, to express it as an integral of a one-half BPS-type superfield which is a descendant of the one-quarter BPS primary \cite{Bossard:2009mn}. To do this, recall that the field strength superfield for MSYM in $D=4, N=4$ is a scalar $W_r$ in the $6$ of $SO(6)$. From this scalar one can construct two scalar bilinears, the Konishi multiplet, $K:=\tr(W_r W_r)$, and the supercurrent $J_{rs}:=\tr (W_r W_s) - \frac{1}{6} \d_{rs} K$. The supercurrent is an ultra-short one-half BPS multiplet; it has $128+128$ components and has an expansion that terminates at fourth order in the odd coordinates. It can be integrated over four odd coordinates and gives rise to the on-shell action \cite{Howe:1981xy}. There are several independent scalar multiplets that can be formed from the square of the supercurrent: the symmetric traceless $105$ is one-half BPS and gives rise to the $\tr^2 F^4$ invariant, while the $84$, which has the symmetries of the Weyl tensor in six dimensions, is the one-quarter BPS multiplet that gives rise to $d^2\tr^2 F^4$. The pseudo-one-half BPS superfield is obtained by inserting two contracted spacetime derivatives, one on each factor,  into the one-half BPS product of two supercurrents. It is a descendant of the one-quarter BPS superfield up to a total spacetime derivative.

A similar situation obtains for the short multiplets in dimensions 5 and 6 \cite{Bossard:2009mn}. In $D\leq 9$ the field strength superfield is a scalar, $W_r,\ r=1\ldots n$ where $n=10-D$, but in $D=7,8$ there are only one-half BPS multiplets and no one-quarter BPS ones because the latter type of constraint is incompatible with manifest Lorentz symmetry in these dimensions.\footnote{Our definition of a BPS superfield is one that is independent of some sets of fermionic coordinates each of which transform under the smallest spinor representation for the spacetime dimension in question. It might be possible to consider splitting up the basic spinor representations, i.e. to use Lorentz harmonics, but we do not do this here.} In $D\leq 8$ the supercurrent is the traceless, symmetric product of two $W$s, and the Konishi multipet is the singlet product. Although there are no one-quarter BPS superfields in $D=7,8$, we can still form pseudo-one-half BPS fields by inserting a pair of contracted spacetime derivatives in the one-half BPS product of two supercurrents, which is again a symmetric, traceless fourth-rank $SO(n)$ tensor.

In dimension 9, the field strength is a singlet $W$ with $\tr W^2$ being the Konishi superfield. The supercurrent is an antisymmetric Lorentz tensor $J_{ab}$ that has dimension three (in units where the dimension of $W$ is one), and has 128 + 128 components. However, in $D=10$, there are no physical scalar fields, and the supercurrent is a dimension-three, third-rank antisymmetric tensor $J_{abc}$ \cite{Howe:1982mt}. It combines the components of both the 128 + 128 and Konishi multiplets and is not locally reducible into a the sum of the two. It is therefore not so easy to generalise integral invariants of the BPS type to $D=9,10$.

The way round this problem is to make use of the so-called ``ectoplasm'' formalism \cite{Voronov,Gates:1997kr,Gates:1997ag}. This allows one to construct spacetime supersymmetric invariant integrals in $D$ dimensions starting from closed $D$-forms in the corresponding superspaces. The idea is to integrate the purely even part of the super-$D$-form, evaluated setting the odd coordinates $\th=0$, over spacetime. The superinvariance of the resulting integral is guaranteed by the fact that the superform, $L_D$ say, is closed. Note that any exact $D$-form will integrate to zero so that we are really concerned with the $D$th cohomology class of super-forms. 

In the next section we shall study supersymmetric invariants in $D=10$. It is well-known that the action itself and the $F^4$ invariants are Chern-Simons invariants, in the sense that they can be constructed, using the ectoplasm formalism, from closed, Weil-trivial $(D+1)$-forms built from invariant polynomials \cite{Berkovits:2008qw}. A feature of the non-abelian  $F^4$ invariants, not previously remarked upon, is that, unless one wishes to introduce explicit dependence on the fermionic coordinates,  the corresponding super-ten-forms have a more complicated structure than those that have been  constructed hitherto. We can also use this formalism to construct the  $d^2\tr^2 F^4$ invariant in $D=10$ by the insertion of two contracted spacetime derivatives. In this case, however, it turns out that the Chern-Simons nature is only apparent and that there is a closed, strictly gauge-invariant superform. Since any $D=10$ invariant gives rise to a $D=4, N=4$ invariant with manifest $SU(4)$ symmetry by dimensional reduction it follows that there cannot be more invariants in $D=10$ than in $D=4$, and this explicit construction shows that there are the same number of short invariants in $D=10$ and $D=4$. A consequence of the existence of the $d^2\tr^2 F^4$ in $D=10$ is that such an invariant exists and has maximal R-symmetry in all dimensions $D\leq 9$; furthermore, as will be discussed in more detail in section four, it has the same cocycle type as the closed super-form for the action, and so is not protected by algebraic non-renormalisation theorems. In addition to the short invariants, section two also contains a discussion of various other features of $D=10$ invariants, such as Konishi and $d^6 F^4$. In section three, we explicitly show how to reduce invariants to lower-dimensional spacetimes and in section five we state our conclusions.


\section{$D=10$ invariants}



\subsection{Basic formalism}


Our conventions are as follows: spacetime indices are $a,b,\ldots$, running from $0$ to $D-1$, while spinor indices are $\a,\b\ldots$ running from 1 to 16 and internal vector indices are $r,s\ldots$, running from $1$ to $n=10-D$; ten-dimensional $16\xz 16$ gamma-matrices are denoted by $\C^a$ and  the even (odd) coordinates of $D=10$ superspace are $(x^a,\th^\a)$. The supersymmetric invariant basis forms are

\bea
 E^a=dx^a-\frac{i}{2} d\th^\a (\C^a)_{\a\b} \th^\b;\qquad  E^\a=d\th^\a\ ,
 \la{2.0}
\eea

and the dual invariant derivatives are $\del_a$ and

\be
 D_\a=\del_\a +\frac{i}{2} (\C^a\th)_\a \del_a\ .
 \la{2.1}
\ee

With respect to this invariant basis an $n$-form splits into a sum of $(p,q)$-forms where $p+q=n$ and $p\,(q)$ denotes the number of even (odd) indices. The exterior derivative splits into three parts, $d=d_0+d_1 +t_0$ with bi-degrees $(1,0),(0,1)$ and $(-1,2)$ respectively. The operation $t_0$, which is purely algebraic, converts a $(p,q)$-form into a $(p-1,q+2)$-form by the contraction of one of the even indices of the form with the vector index of the dimension-zero torsion $T_{\a\b}{}^c=-i(\C^c)_{\a\b}$ and then by the symmetrisation of all $(q+2)$ odd indices on the form and the torsion. It is not difficult to see that $t_0^2=0$, so that there are associated cohomology groups $H_t^{p,q}$ \cite{Bonora:1986ix}. In $D=10$ these vanish for $p>6$ while the cohomology for $p\geq 1$ is related to the five-index gamma matrix except for two additional contributions in $H_t^{1,1}$ and $H_t^{1,2}$ whose existence is due to the single-index gamma-matrix \cite{Berkovits:2008qw}. For $p=0$, $H_t^{0,q}$ is isomorphic to the space of pure $q$-spinors, i.e. symmetric, gamma-traceless $(0,q)$-forms. We can also define an odd derivative that acts on $t_0$-cohomology. It is defined by $d_s[\o_{p,q}]=[d_1\o_{p,q}]$, where the brackets denote $t_0$-cohomology classes. It is easy to check that the spinorial derivative $d_s$ is well-defined and that it squares to zero so that we can define the so-called spinorial cohomology groups $H_s^{p,q}$ \cite{Cederwall:2001dx,Howe:2003cy}. It is a generalisation of pure spinor cohomology with which it coincides in the case $p=0$ \cite{Berkovits:2002zk}.

The field strength two-form in $D=10$ superspace has $F_{\a\b}=0$ and $F_{a\b}=(\C_a)_{\b\c}\L^\c$. The Lie-algebra valued field strength superfield $\L^\a$ has the spinor field of the $D=10$ SYM multiplet as its leading component; it obeys the constraint

\be
\nab_\a \L^\b=-\frac{i}{4}(\C^{ab})_\a{}^\b F_{ab}\ ,
\la{2.3}
\ee

where $F_{ab}$ is the $(2,0)$ component of the field strength whose leading component is the spacetime Yang-Mills fields strength. The form of $F_{a\b}$ and \eq{2.3} follow from the basic constraint $F_{\a\b}=0$ which can be viewed as a pure spinor integrability condition \cite{Howe:1991mf,Howe:1991bx}. By use of the Bianchi identity one can then show that the field equations hold for $\L^\a$ and $F_{ab}$. They are

\bea
\C^a \nab_a \L&=& 0 \nn\w1
\nab^b F_{ab}&=&2i \L\C_a\L\ ,
\la{2.4}
\eea

where the right-hand side is Lie-algebra valued because the quadratic expression in $\Lambda$ forces antisymmetry in the group indices. 

The supercurrent is

\be
J_{abc}=\tr (\L\C_{abc}\L)\ .
\la{2.5}
\ee

The simplest full superspace integral of a gauge-invariant integrand that can be constructed in $D=10$ is the integral of the square of the supercurrent, which leads to a spacetime integral of $d^6\tr^2 F^4$. (The double-trace is required because the symmetrised single-trace of $\L^4$ vanishes identically.\footnote{There are other $\L^4$ possibilities involving $\L\C_1\L$ or $\L\C_5\L$ but these contain gauge commutators.}) Furthermore, there are no gauge-invariant BPS superfields that one can construct. This means that the construction of invariants with lower dimensionality requires a different approach to $D\leq 8$. As we mentioned in the introduction, this is the ectoplasm formalism. In $D$ spacetime dimensions, given a closed super $D$-form, $L_D$, in superspace, the integral

\be
I=\int d^D x\,\ve^{m_1\ldots m_D} L_{m_1\ldots m_D} (x,\th=0)
\la{2.6}
\ee

is supersymmetric owing to the fact that $L_D$ is closed. Now suppose we are given a closed super $(D+1)$-form $W_{D+1}=d Z_D$, where $Z_D$ is an explicit potential form that is not gauge-invariant, and such that $W_{D+1}$ is Weil-trivial \cite{Bonora:1986xd}, i.e. $W_{D+1}=d K_D$, where $K_{D}$ is gauge-invariant, then $L_D= K_D- Z_D$ is closed and so can be integrated using the ectoplasm formula. Such an invariant is called a Chern-Simons invariant owing to the presence of the Chern-Simons form $Z_D$.  Green-Schwarz actions for branes are examples of this \cite{Howe:1998tsa}. 

In order for a closed $D$-form  to give rise to an invariant it must have a non-vanishing $L_{D,0}$ component, at least in flat superspace which is the case of interest here. It will have a lowest non-zero component, $L_{p,q}$ say, (lowest means lowest mass dimension, i.e. smallest $p$) that must satisfy
$t_0 L_{p,q}=0$ and that must not be of the form $t_0 K_{p+1,q-2}$ since such a term vanishes in $t_0$ cohomology. So the lowest component will be defined by an element of $H_t^{p,q}$, the cohomology associated with the nilpotent operator $t_0$, and moreover it must be $d_s$-closed, i.e. it must be an element of $H_s^{p,q}$. In $D=10$ $H_t^{p,q}$ vanishes for $p>5$ so that the simplest possibility for the lowest non-vanishing component of a closed super-ten-form is an $L_{5,5}$  that   must have the form $L_{5,5}=\C_{5,2} M_{0,3}$ where $d_s[M_{0,3}]=0$. The integral invariant corresponding to such a form can be expressed as a Berkovits superaction  integral over five thetas \cite{Berkovits:2002zk}. The spacetime integrand of such an invariant is $[D^5]^{\a\b\c}\, M_{\a\b\c}$; it involves the contraction of two $42\cdot 16$-dimensional representations with opposite chirality, $[00030]$ and $[00003]$ in terms of Dynkin labels.

However, there can be other more general types of closed super-ten-forms as we shall discuss below. Indeed, more generally,  we can ask the question of when a $d_s$-closed element $[\o_{p,q}]$ of $H_t^{p,q}$ determines a closed $n$-form, for $n=p+q$. The answer is that there is a sequence of possible obstructions, $[\l_{p+1,q}],[\l_{p+2,q-1}],[\l_{p+3,q-2}],\ldots$  ,each of which is $d_s$ closed provided that the preceding ones in the sequence are $d_s$ exact. If they are all exact, and if we set $[\l_{p+r,q-r+1}]=d_s[\m_{p+r,q-r}]$, then the entire closed $n$-form will be determined by the sequence of elements $[\o_{p,q}],[\m_{p+1,q-1}],\dots $. This is easy to see; suppose $\o_{p,q}$ is a representative of $[\o_{p,q}]$, then there is an $\o_{p+1,q-1}$ such that $d_1 \o_{p,q}+t_0\o_{p+1,q-1}=0$. Hitting this  equation  with $d_1$ we find that $d_0\o_{p,q}+ d_1\o_{p+1,q-1}:=\l_{p+1,q}$ is $t_0$-closed although not necessarily exact. It is also easy to check that the associated $t_0$-cohomology class,  $[\l_{p+1,q}]$,  is $d_s$-closed. If this is $d_s$ exact, we can redefine $\o_{p+1,q-1}$ and deduce the existence of an $\o_{p+2,q-2}$ such that $d_0\o_{p,q} + d_1\o_{p+1,q-1}+ t_0 \o_{p+2,q-2}=0$, which is the $(p+1,q)$ component of the equation for a closed
$(p+q)$-form. We can then iterate this procedure by applying $d_1$ to this equation.  

For $D=10$  the $t_0$-cohomology groups $H_t^{p,q}$ can be related (via the five-index $\C$-matrix) to cohomology groups for pure spinor $(q-2)$-forms taking their values in $\L^{5-p} T_0$, where $T_0$ is the even tangent bundle, i.e. tensors with $k=(5-p)$ antisymmetrised vector indices in addition to $(q-2)$ spinorial form indices, modulo equivalences.\footnote{There are two exceptional cases: $H_t^{1,1}$ and $H_t^{1,2}$, but they are not relevant to the forms we are interested in here.} We denote these cohomology groups by $H_t^{0,q-2}(\L^k T_0)$. It is also possible to extend the definition of $d_s$ to act on these groups \cite{Berkovits:2008qw}, so  the obstructions to a $d_s$-closed $(p,10-p)$-form determining a closed super-ten-form lie in the groups $H_s^{n,11-n}\cong H_s^{0,9-n}(\L^{5-n} T_0)$, for $n=p+1,\ldots 5$. Clearly $0\leq p\leq 5$, while the last obstruction will lie in $H_s^{0,4}$. If the lowest component of the ten-form lies in $H_s^{5,5}$ there are no obstructions (this is the usual case), but if the lowest component has $p<5$ there can be obstructions. This turns out to be the case for the Chern-Simons $F^4$ invariants as we shall see below.


\subsection{Chern-Simons invariants}


The $D=10$ Chern-Simons invariants are constructed from the invariant polynomials $P_4:= \tr F^2$, $P_8:=\tr F^4$ and $P'_8:= (\tr F^2)^2$. They are written in terms of Chern-Simons forms as $P_4= dQ_3$, $P_8=d Q_7$ and $P'_8= dQ'_7$. Clearly a possible choice for $Q'_7$ is $Q'_7= Q_3 P_4$.  The on-shell action is itself of this type; the closed $11$-form is

\be
W_{11}=H_7 P_4 \ ,
\la{2.7}
\ee

where $dH_7=0$ and in flat superspace $H_7$  is proportional to $\C_{5,2}$. It is easy to see that $W_{11}$ is Weil trivial and that the lowest component of the corresponding $K$ is 
 \be K_{8,2} \sim  i  \C_{5,2} J_{3,0} + 3 i \C_{1,2} \star J_{3,0} \; \; .\ee
To see this, note that the lowest component of $P_4$ is $P_{2,2}$ with

\be
P_{2,2}\sim i t_0 J_{3,0} +  \frac{1}{6} \C_{2abc,2}J^{abc} \  ,
\la{2.8}
\ee

We can choose $Z_{10}=H_7 Q_3$ so that the lowest non-vanishing component of $L$ is clearly $L_{5,5}=-Z_{5,5}$. In fact, one simply has $M_{0,3}\sim Q_{0,3}$. 

The $F^4 $ invariants involve $P_8$ or $P'_8$ multiplied by the closed three-form $H_3$ which in flat superspace is proportional to $\C_{1,2}$. Again it is not difficult to prove Weil triviality \cite{Berkovits:2008qw}. In both of these cases the lowest component of the corresponding $K$ is $K_{6,4}$. For the double-trace,
 
 \be K_{6,4} = i  \C_{1,2} J_{3,0} P_{2,2} + \frac{i}{2} \C_{3ab,2}  J^{ab}{}_{1,0} P_{2,2}  \; \; , \label{K64}  \ee

and a similar analysis applies in the single-trace case with $K_{6,4} \sim \tr \L^4$. 

At this stage we have two options for $Z_{10}$. The simpler one is to keep manifest Yang-Mills gauge invariance and to choose a gauge for the `external' two-form $B$, the potential of $H_3$. This involves the introduction of an explicit factor of $\th$ into the problem, but its simplicity makes it useful for other purposes. However, in order to understand the structure of the cocycle relevant to non-renormalisation theorems within the framework of algebraic renormalisation, it is important to consider a cocycle which does not depend explicitly on $\th$, for which the components are related by supersymmetry  in the normal fashion. This requires  the introduction of the Yang-Mills Chern-Simons seven-forms without specifying any particular gauge, {\it i.e.} to choose the Wess-Zumino parts of the closed super-ten-forms to be $Z_{10} = H_3 Q_7  $ and $Z_{10} = H_3ÊQ_3 P_4$, respectively.  It turns out that the closed super-ten-forms in this second approach involve non-vanishing lowest components with even degree less than five, whereas the former method leads to forms that start at $L_{5,5}$.

\subsubsection*{Chern-Simons cocycles}

Consider first the double-trace $F^4$ invariant in this second approach. The lowest component $Z_{3,7} \sim \C_{1,2}ÊQ_{0,3}  P_{2,2}$ is $t_0$-trivial, so that we can find a cohomologically equivalent $Z'$ which has lowest component

\be 
Z^\prime_{4,6} \sim \C_{4a,2} N^a{}_{,4} \; ,  
\la{4.1}
\ee

where $N_{1,4} \sim \hat{\Xi}_{1,1} \hat{Q}_{0,3}$ is projected into the $[10004]$ component.\footnote{The notation in this equation signifies that one of the even indices on $\C_{5,2}$, labelled by $a$, is contracted with the even vector index on $N$. This notation is used in the following for both even and odd indices. } The lowest component of the ten-form is therefore $L_{4,6}=-Z'_{4,6}$. From the discussion at the end of subsection 2.1 we see that $N$ defines an element of $H_t^{0,4}(T_0)$ which must be $d_s$-closed. This in turn leads to a possible obstruction to the existence of a closed ten-form with this lowest component. It lies in $H_s^{5,6}\cong H_s^{0,4}$ and must be trivial because $L_{10}$ is closed. We therefore have a $(0,4)$-form, $O_{0,4}$, such that $[O_{0,4}]=d_s[M_{0,3}]$ for some $M_{0,3}$. If the form $O_{0,4}$ were to be zero, then the cocycle  derived directly from $L_{4,6}$ would have a vanishing top component, $L_{10,0}$, because  $D^6 N_{1,4}Ê$ does not contain a singlet. Furthermore, there would have to be a $d_s$-closed $M_{0,3}$ in order to obtain a non-zero top component.  In other words, the cocycle would split into two irreducible multiplets, one with a standard $L^{\rm \scriptscriptstyle inv}_{5,5}$ first component defining the invariant, and the other starting as $L^{\rm \scriptscriptstyle cur}_{4,6}$. The highest non-zero component of $L^{\rm \scriptscriptstyle cur}$, say $L^{\rm \scriptscriptstyle cur}_{9,1}$, would define a conserved current  (in spacetime)  which would not be a total derivative (\ie its divergence would only vanish because of the equations of motion). However, there is no such conserved current within the multiplet of $N_{1,4}$. It therefore follows that $O_{0,4}$ cannot vanish. Furthermore, there can be no independent $d_s$-closed $[M_{0,3}]$ because there is only one invariant. In the following we shall confirm these expectations explicitly.

The [10004] projection of $N$ is given by

\bea N^a{}_{,4}Ê&=&\hat{\Xi}^a{}_{,1} \hat{Q}_{0,3} + \frac{7}{96} \C^b{}_{,2} \C_b^{\a\b} Ê\hat{\Xi}^a_{\a} \hat{Q}_{0,2\b} +  \frac{1}{96} \C_{b,2} \C^{a,\a\b} Ê\hat{\Xi}^b_{\a} \hat{Q}_{0,2\b} +  \frac{1}{16}  [\C^a \C_b]_{,1}{}^\a  \hat{\Xi}^b{}_{,1} \hat{Q}_{0,2\a}  \nn  \w1 
& & \hspace{10mm}Ê+  \frac{1}{192} [\C^a \C_b]_{,1}{}^\a \C^c{}_{,2} \C_{c,}{}^{\b\c}Ê\hat{\Xi}^b_{\c} \hat{Q}_{0,1\a\b} \; \; , \eea

with $ \hat{\Xi}^a$ the gamma-traceless component of the supersymmetry current 

\be 
\hat{\Xi}^a \equiv  \frac{i}{10} \tr \Bigl(Ê\Bigl[ -8 F^{ab} \C_{b} +  F_{bc}Ê\C^{abc} \Bigr] \Lambda \Bigr) \; ,
\la{4.2} 
\ee 

and $\hat{Q}_{0,3}$ the $[00003]$ component of $Q_{0,3}$

\be \hat{Q}_{\a\b\c} \equiv Q_{\a\b\c} - \frac{3}{20}Ê\C^a_{(\a\b} \C_a^{\d\h}ÊQ_{\c)\d\h} \; \; . \ee

 $Z^\prime_{4,6}$ is not $t_0$-trivial, and the super-ten-form associated to the double-trace $F^4$ invariant thus does not admit a representative with a vanishing $L_{4,6}$.

 A possible  obstruction to the definition of the $(5,5)$ component  would be  associated to a $(4,7)$ $t_0$-cohomology class

\be  d_1 Z^\prime_{4,6}(N) \sim \C_{4a,2} O^{a}{}_{,5}(N) \ee

where $O^{1}{}_{,5}$  would be   an operator in the $[10005]$. However, $D N_{1,4}$ vanishes in the $[10005]$, and  so there is no such obstruction. The obstruction to the definition of the $(6,4)$ component is associated to a $(5,6)$ $t_0$-cohomology class given by

\be  d_0 Z^\prime_{4,6}(N)  + d_1 Z^\prime_{5,5}(N)   \sim \C_{5,2} \,  O_{0,4}(N) \label{Obstruction} \ee

where $O_{0,4}$ is an operator in the $[00004]$. There are two distinct components in this representation descending from $N_{1,4}$, so that

\be [O_{0,4}] \sim [D^2]_{abc}\,  [\C^{ab}]_{1}{}^\a  N^c{}_{,3\a}   + c \,  \partial_a N^a{}_{,4}  \ee

for some  coefficient $c$ determined such that $[d_1 O_{0,4}] = 0$. This is possible because the only component of $D^3 N_{1,4}$ in the $[00005]$ is the total derivative 

\be [ d_1   \partial_a N^a{}_{,4} ] \sim \C^5{}_{,2} \hat{P}_{4,0} d_0 \hat{Q}_{0,3} \; , \ee 

where~\footnote{ This differs slightly from  the component that occurs in the superconformal multiplet.}Ê
\be \hat{P}_{abcd} \equiv  \trace  F_{[ab} F_{cd]}  - \frac{i}{2}Ê\partial_{[a}ÊJ_{bcd]} \ .  \ee

To discuss $O_{0,4}$ more explicitly, let us describe some relevant components of the superfield $\hat{Q}_{0,3}$. $D \hat{Q}_{0,3}$  decomposes into two irreducible components of $Q_{1,2}$ and $S_{abc,\a\b} \equiv \frac{1}{16} \C_{abc}^{\c\d} D_\c Q_{\a\b\d} $, as
\be D_\d \hat{Q}_{\a\b\c}Ê= \frac{15i}{8} \C^a_{\d(\a} \hat{Q}^\prime_{a,\b\c)} - \frac{3i}{8} \C^a_{(\a\b} \hat{Q}^\prime_{a,\c)\d} + \frac{1}{4} \C^{abc}_{\d(\a}Ê\hat{S}_{abc,\b\c)}Ê\; \; ; \ee
where the $[10002]$ 
\bea \hat{Q}^\prime_{1,2} &\equiv& \frac{2}{5} \left( Q_{1,2} + \frac{1}{96} \C_{1,2} \C^{a,\a\b} Q_{a,\a\b} - \frac{1}{96} \C^a{}_{,2} \C_{1,}{}^{\a\b} Q_{a,\a\b} - \frac{5}{96}  \C_{a,2} \C^{a,\a\b} Q_{1,\a\b}  \right . \nn\w1&&\quad \qquad  \left . - \frac{1}{12} [\C_1 \C^a]_{,1}{}^\a Q_{a,1\a} \right) 
 - \frac{i}{20} \left( \C^{ab}{}_{,1}{}^\a S_{1ab,1\a} + \frac{1}{6} \C^b{}_{,2} \C^{a,\a\b} S_{1ab,\a\b}  \right . \nn\w1&&\qquad\qquad  \left .  - \frac{1}{12} [\C_1 \C^{abc}]_{,1}{}^\a S_{abc,1\a} + \frac{1}{12} \C^{ab}{}_{,1}{}^\a [\C_1 \C^c]_{,1}{}^\b S_{abc,\a\b} \right)
\eea
and the $[00102]$ $\hat{S}_{3,2}$ is the corresponding component of $S_{3,2}$. $D^2 \hat{Q}_{0,3}$  decomposes similarly into  $\hat{Q}^\prime_{2,1}$ in the $[01001]$, $\hat{S}_{abcd,\a}$ in the $[00012]$, $\hat{S}_{abc,d,\a}$ in the $[10101]$ and $\hat{S}_{abcd,ef,\a}$ in the $[01012]$. 

 Using representation theory,  one computes  that (where $\Xi^\a \equiv \C^{ab,\a}{}_\b  \tr F_{ab} \Lambda^\b$)

\bea \bigl[ÊO_{0,4} \bigr]Ê&\sim& c_1 \hat{\Xi}^a{}_{,1} \partial_a \hat{Q}_{0,3}Ê+ c_2 \partial_a \hat{\Xi}^a{}_{,1} \hat{Q}_{0,3}+ c_3  [\C^{ab}]_1{}^\a \partial_a \hat{\Xi}_{b,1} \hat{Q}_{0,2\a} + c_4  [\C^{ab}]_{1}{}^\a \partial^c ( d_1 J_{abc}Ê) \, \hat{Q}_{0,2\a} \nn\w1  
& &  + \C^{abcde}{}_{,2}Ê\Bigl( c_5  \hat{P}_{abcd} \hat{Q}^\prime_{e,2}Ê + c_6 \partial^f J_{abf}Ê\hat{S}_{cde,2} + c_7  \partial^f J_{abc}Ê\hat{S}_{def,2} + c_8 
\hat{P}_{abc}{}^f  \hat{S}_{def,2}Ê+ c_9 \hat{\Xi}_{a,1}Ê\hat{S}_{bcde,1}   \Bigr . \nn \w1  && \hspace{20mm}  \Bigl . Ê+ c_{10}  [\C^{fg}]_{1}{}^{\a}Ê\hat{\Xi}_{a\a}Ê\hat{S}_{bcde,fg,1} \Bigr)  +  c_{11}  [\C^{ab}]_1{}^\a \hat{\Xi}_{a,1} \partial_b \hat{Q}_{0,2\a}Ê+  c_{12} [\C^a]_{1\a}Ê\Xi^\a   \partial_a \hat{Q}_{0,3}  \nn\w1
&\sim&  - d_1 M_{0,3} \; , \label{Otrivial}\eea

with

\bea M_{0,3} &\sim& c^\prime_1 T Q_{0,3} + c^\prime_2 [\C^{ab}]_{1}{}^\a \partial^c  J_{abc}Ê \, \hat{Q}_{0,2\a} + c^\prime_3  [\C^{abcd}]_{1}{}^\a \hat{P}_{abcd}  \hat{Q}_{0,2\a} +c^\prime_4 \hat{\Xi}^a{}_{,1}ÊQ_{a,2} \hspace{10mm}  \nn\w1 & & \hspace{3mm}  + c^\prime_5 \hat{\Xi}^a{}_{,1}Ê\hat{Q}^\prime_{a,2} +  c^\prime_6 \C^a{}_{,1\a} \Xi^\aÊ\hat{Q}^\prime_{a,2} +  c^\prime_7 [\C^{ab}]_1{}^\a \hat{\Xi}^c_\a \hat{S}_{abc,2} + c^\prime_8  [\C^{abc}]_{1\a}Ê\Xi^\a \hat{S}_{abc,2}  \; . \eea

Therefore 

\be Z^\prime_{5,5} =  Z^\prime_{5,5}(N) + \C_{5,2} M_{0,3} \; ,   \label{Ztotal}\ee

but $M_{0,3}$ is clearly not a descendant of $N_{1,4}$ since $D N_{1,4}$ does not  have  a $[00003]$ component. Moreover,  two  terms  in  $M_{0,3}$ involve the trace of the energy momentum tensor $T$, which is not a descendant of $\hat{\Xi}^1{}_{,1}$, and the component of $Q_{1,2}$ in the $[00011]$ which is not a descendant of $\hat{Q}_{0,3}$. To  show  that $O_{0,4}$ indeed appears with a non-vanishing coefficient, it is enough to compute that one coefficient does not vanish, say $c_{10}$.  We have 

\be \C_{4a,2} ÊN^a{}_{, 4} = \C_{4a,2} Ê\hat{\Xi}^a{}_{, 1} \hat{Q}_{0,3} + t_0 V_{5,4} \ee

with 

\bea V_{5,4}Ê&\equiv& \frac{7i}{96}Ê\C_{4a,2} \C_{1,}{}^{\a\b}Ê\hat{\Xi}_\a^a \hat{Q}_{0,2\b}+  \frac{i}{96}Ê\C_{4a,2} \C^{a,\a\b}Ê\hat{\Xi}_{1,\a} \hat{Q}_{0,2\b}Ê
Ê- \frac{i}{16}Ê[ \C_5 \C_a]_{,1}{}^\a \hat{\Xi}^a{}_{,1} \hat{Q}_{0,2\a} \nn\w1 & & \hspace{30mm} Ê- \frac{i}{192}ÊÊ\C_{4a,2} [\C^a \C_b]_{,1}{}^\a \C_{1,}{}^{\b\c}Ê \hat{\Xi}_\c^b \hat{Q}_{0,1\a\b} \; \; . \eea

One computes 
\bea  d_1\left(\C_{4a,2}  ÊN^a{}_{, 4}\right)  &=& - t_0 \biggl( \C_{4a,2}  \hat{\Xi}^a{}_{,1}Ê\hat{Q}^\prime_{1,2}Ê - i \C_{4a,2}  \Bigl(  \hat{T}_{1}{}^a  + \frac{i}{5}Ê\partial_b J_1{}^{ab} \Bigr) \hat{Q}_{0,3}
\biggr .  \nn \w1
& & \hspace{20mm}  \left .Ê- \frac{6i}{5} \C_{1,2} \hat{P}_4 \hat{Q}_{0,3}Ê+ \frac{3i}{5}Ê\C_3{}^{ab}{}_{,2}Ê\hat{P}_{2ab} \hat{Q}_{0,3}Ê+ d_1 V_{5,4}Ê \right) \; ,  \eea

and 

\bea d_0\left(\C_{4a,2}  ÊN^a{}_{, 4}\right) &+&  d_1 \left( \C_{4a,2}Ê\hat{\Xi}^a{}_{,1}Ê\hat{Q}^\prime_{1,2}Ê+ \cdots + d_1 V_{5,4}Ê\right) \nn\w1&&  + t_0 \left( - i \Bigl[\C_{4a,2} \Bigl( \hat{T}_1{}^a  + \frac{i}{5}Ê\partial_b J_1{}^{ab} \Bigr)  + \frac{6}{5} \C_{1,2}   \hat{P}_{4}  - \frac{3}{5}Ê\C_3{}^{ab}{}_{,2}Ê\hat{P}_{2ab} \Bigr] \hat{Q}^\prime_{1,2} + d_0 V_{5,4}Ê \right)  \nn\w1 &=&  \C_{4a,2} \Bigl( \hat{\Xi}^a{}_{,1} ( d_0 \hat{Q}_{0,3} + d_1 \hat{Q}^\prime_{1,2} ) - d_0 \hat{\Xi}^a{}_{,1}  \hat{Q}_{0,3}\Bigr) \nn\w1 && \qquad + i d_1 \Bigl[\C_{4a,2} \Bigl( \hat{T}_1{}^a  + \frac{i}{5}Ê\partial_b J_1{}^{ab} \Bigr)  + \frac{6}{5} \C_{1,2}   \hat{P}_{4}  - \frac{3}{5}Ê\C_3{}^{ab}{}_{,2}Ê\hat{P}_{2ab} \Bigr] \hat{Q}_{0,3} \; . \label{DescentDeb}\eea 

$O_{0,4}$ does not appear here explicitly, but one must take into account that the $ Z^\prime_{5,5}$ representative that we have used  involves the components of $\hat{\Xi}^a{}_{,1} \hat{Q'}_{a,2}$ and $\C^{abcd}{}_{,1}{}^\a \hat{P}_{abcd} Q_{0,2\a}$ in the $[00003]$ which do not appear in $D N_{1,4}$. One computes in the same way that $d_1 V_{5,4}$ involves  a  non-vanishing $ [\C^{ab}]_1{}^\a \hat{\Xi}^c_\a \hat{S}_{abc,2}$ term in the $[00003]$, exhibiting that $c^\prime_7= -\frac{11i}{1152}$. Because 
\be d_1  [\C^{ab}]_{,1}{}^\a \hat{\Xi}^c_\a \hat{S}_{abc,2} \sim  \C^{abcde}{}_{,2}\,  \C^{fg}{}_{,1}{}^\alpha \hat{\Xi}_{a,\alpha}Ê\hat{S}_{bcde,fg,1} + \cdots \ee
in pure spinor cohomology, and because such a term can only be cancelled by the corresponding terms in ${d_1} V_{5,4}$ involving the components of $\hat{\Xi}_{1,1} \hat{S}_{3,2}$ in the $[20003]$, the $[01003]$, the $[00014]$ and the $[10103]$ which appear in the corresponding components of $D N_{1,4}$, we conclude that $c_{10}$  is also non-zero. In particular, as exhibited in 
(\ref{DescentDeb}), the unprojected $ \C_{4a,2} Ê\hat{\Xi}^a{}_{, 1} \hat{Q}_{0,3}$ only contribute to  $O_{0,4}$ through terms in  $D^2 \hat{\Xi}_{1,1} \hat{Q}_{0,3}$, $\hat{\Xi}_{1,1} d_0 \hat{Q}_{0,3}$ and $\hat{\Xi}_{1,1} d_1 \hat{Q}^\prime_{1,2}$, and therefore only contribute to the coefficients $(c_2,\, c_3,\, c_4)$, $(c_1, \, c_{11})$ and $c_{9}$, respectively. 

 $M_{0,3}$  is not a supersymmetry descendant of $N_{1,4}$, but $O_{0,4}$  is a supersymmetry descendant of both $M_{0,3}$ and $N_{1,4}$, and thereby provides the `bridge' which relates the double-trace invariant to $N_{1,4}$. The field $M_{0,3}$ satisfies the constraint 
 
 \be d_1 M_{0,3} \sim -  [D^2]_{abc}\,  [\C^{ab}]_{1}{}^\a  N^c{}_{,3\a}   -  c \,  \partial_a N^a{}_{,4}  \ee
 
 in pure spinor cohomology, which is weaker than the conventional linear constraint, but strong enough in order for 
 $[D^5]^{\a\b\c} M_{\a\b\c}$ to define a supersymmetry invariant in spacetime. To see this one makes a supersymmetry variation, {\it i.e.} one applies another $D$ to $[D^5]^{\a\b\c} M_{\a\b\c}$, and verifies that the result can only be a spacetime divergence. This makes use of the fact that the component of $D^7 N_{1,4} $ in the $[00001]$ is a total derivative by representation theory.
 
Note that this implies that this cocycle differs in structure from a cocycle associated to an ordinary superspace integral in another sense, namely that, although $L_{9,1}$ usually only involves a $[00010]$, this cocycle has, in addition, a  component in the $[10001]$ representation {\it i.e.}  
\be L_{9,1} \sim  \C_{9,1\a} [D^4]^{\a\b\c\d}  M_{\b\c\d} + [D^5]^{\a\b\c}ÊN_{9,1\a\b\c} Ê\; \; . \label{L91}Ê\ee

The discussion of the single-trace invariant is rather similar. In this case, the lowest component 

\be 
Z_{1,9} \sim \C_{1,2} \, Q_{0,7} \; ,
\la{4.7}
\ee

is again $t_0$-trivial, so that one can choose a cohomologically equivalent representative $Z^\prime$ such that its lowest component is
 
\be 
Z^\prime_{2,8} \sim \frac{1}6{} \C_{2abc,2} \hat{S}^{abc}{}_{,6}   \; , 
\la{4.8} 
\ee

where $S_{3,6} \equiv [\C_{3}]^{\a\b} D_\a Q_{0,6\b}$ projected into the $[00106]$. Again $Z^\prime_{2,8}$ is neither  $t_0$-trivial nor $d_1$-exact, and there is no super-ten-form representative of the invariant with a vanishing $(2,8)$ component. The supermultiplet structure of the super-ten-form is similar to the one of the double-trace invariant, although one must encounter at least two `bridges' as in (\ref{Obstruction},\ref{Otrivial},\ref{Ztotal}), instead of just one, in order to obtain the last component $Z^\prime_{2,8}$ from the invariant $L_{10,0}$.\footnote{One `bridge' is necessarily associated to a $(5,6)$  $t_0$ cohomology class, and the second could be associated to either a $(3,8)$ or a $(4,6)$  $t_0$ cohomology class, or all three of them could be necessary.}

\subsubsection*{$B$-field cocycles}

As discussed in the beginning of this section, one can also remove the lowest component in order to obtain a gauge invariant cocycle starting as $L_{5,5} \sim \C_{5,2} M_{0,3}$, at the cost of introducing an explicit dependence on the fermionic coordinates $\th$. In order to do this we shall take $Z_{10} = B_2 P_8$ where $dB_2=H_3$. In flat superspace one can choose $B_{1,1}=\i_{\th} \C_{1,2}$ where $\i_{\th}$ indicates the contraction of a form with the odd vector field $\th^\a D_\a$. After some gamma-matrix algebra, one can show that the double trace $L_{5,5} \sim B_{1,1} P_{2,2} P_{2,2}$ is equivalent (up to $t_0$-exact terms) to an expression of the standard form, $\C_{5,2} M_{0,3}$,  with

\be
M_{0,3}\sim \i_\th\C_{a,2} Y^a{}_{,2}\ 
\la{2.12}
\ee

where 

\be
Y_{a\b\c}=(\C^{bcdef})_{\b\c} J_{abc} J_{def}\ .
\la{2.13}
\ee

In this expression, we can take  the product of the two supercurrents to be in the 
 [10002]  
 representation of the Lorentz group owing to the self-duality of the gamma-matrix and the fact that the four-form trace in $J^2$ gives a $t_0$-exact term in \eq{2.13}. It is not difficult to see that  $L_{5,5} \sim B_{1,1} P_{4,4}$ has the correct form in the single trace case as well.

Any pure spinor integral can be trivially rewritten as

\be
 \int d^{10} x\, [D^{5}]^{\a\b\c}\, M_{\a\b\c} =\int d^{10}x\, [D^4]^{a\a\b} [DM]_{a\a\b}\ ,
\ee

where the integrand in the second expression is in the [10002] representation, but no longer obeys a simple linear constraint in $D$. If we do this for the double-trace invariant, we can get rid of the explicit $\th$ and rewrite it in the form

\be
I=\int d^{10} x  [D^4]^{a\a\b} \, Y_{a\a\b} \ ,
\ee

where $Y_{1,2}$ is given by \eq{2.13}.


\subsection{The $d^2 \tr^2 F^4$ invariant}



We can also use the double-trace Chern-Simons invariant to construct the $d^2\tr^2 F^4$ invariant in $D=10$. This can be done by inserting two contracted spacetime derivatives into the eleven-form $H_3 P'_8= H_3 P_4 P_4$. We can write this as $H_3 \del^a P_4 \del_a P_4$, where the derivative of a form is the form with the derivative acting on its components. The above analysis goes through in exactly the same way in the presence of the derivatives. For $H_3 P_4 P_4$, we have seen in the last section that there is a representative defined by $M_{0,3} \sim \th J^2 $ (\ref{2.13}), so when the derivatives are inserted we shall simply get $\th \del^a J \del_a J$ leading to a spacetime integrand of a similar form to that discussed in the introduction for this type of invariant in lower-dimensional spacetimes. Note that such a procedure cannot be used for the single-trace invariant as the derivatives would either have to sit outside the trace, leading to a total derivative, or, if taken inside, would have to be covariant and hence spoil closure of the $W$-form.

We have therefore succeeded in constructing the double-trace $d^2 F^4$ invariant in $D=10$, but it actually turns out to be expressible in terms of a strictly gauge invariant closed super-ten-form. To see this, note first that $\del_a \o=\cL_a \o$, where $\o$ is a form and $\cL_a$ denotes the Lie derivative along the basis vector $E_a=\del_a$ in flat superspace. We therefore have

\be
\cL_a P= \i_a d P + d \i_a P =d \i_a P\ ,
\la{2.9}
\ee

where $\i_a$ denotes the interior product with the basis vector $E_a$ and $P$ is an invariant polynomial form. Therefore the eleven-form $\hat W_{11}$

\bea
\hat W_{11}:=H_3 \cL_a P_4 \cL^a P_4&=& H_3 (d \i_a P_4)\cL^a P_4\nn\w1
&=& d(H_3 \i_a P_4 \cL^a P_4) := d \hat V_{10}\ ,
\la{2.10}
\eea

since $d\cL_a=\cL_a d$. But we already knew that the left-hand side can be written as $d\hat K_{10}$, for some $\hat K_{10}$ (and also as $d\hat Z_{10}$), so that $L^{\rm \scriptscriptstyle inv}_{10}:=\hat K_{10}-\hat V_{10}$ is closed and strictly gauge-invariant. 

We shall now argue that  $L^{\rm \scriptscriptstyle inv}_{10}$  cannot be exact. To see this, let us go back to the superform $\hat L_{10}=\hat K_{10}-\hat Z_{10}$. Since $\hat Z_{10,0}=0$ it follows that the spacetime integrand is given by $\hat K_{10,0}$. We know that this is not zero because it is simply the same as that for the $\tr^2 F^4$ invariant but with two inserted derivatives. But similarly, we can see that $\hat V_{10,0}=0$ and so the new, gauge-invariant $L^{\rm \scriptscriptstyle inv}_{10}$ also gives rise to a spacetime integral of $\hat K_{10,0}$. If $\hat K_{10}-\hat V_{10}$ were exact, this integral would have to vanish, which it does not, as we have just argued. This means that there is a $d^2\tr^2 F^4$ invariant in $D=10$, and that it can be derived from a strictly gauge-invariant closed superform.

%



 The lowest component of $\hat{K}_{10}$ is $\hat{K}_{6,4}$, which can simply be obtained from (\ref{K64}) by inserting derivatives.  So the lowest component of $L^{\rm \scriptscriptstyle inv}$ is determined by $\hat V$. As defined, this is actually $\hat V_{4,6}$, so we first have to show that this is $t_0$ exact. This turns out to be the case with

\be
\hat V_{4,6}=\C_{1,2} \i_a P_{2,2} \cL^a P_{2,2}=-t_0 N_{5,4}\ ,
\la{2.14}
\ee

where

\be
N_{5,4}=\i_a P_{2,2} X^a_{4,2}\  ,
\ee

and where

\be
\C_{1,2}\cL^a P_{2,2}=t_0 X^a_{4,2}\ .
\la{2.15}
\ee

In fact, a similar equation to \eq{2.15} holds for $P_{2,2}$ without the derivative, so that $X^a\sim \del^a J$. So now, if we add a term $d N$ to $\hat V$, where $N$ has lowest component $N_{5,4}$, the resulting object will have a $(5,5)$ lowest component which is $t_0$ closed, and we just have to rewrite this in the form $\C_{5,2} M^{\rm \scriptscriptstyle inv}_{0,3}$ to find the lowest component of the gauge-invariant superform for $d^2\tr^2 F^4$.  After some algebra one finds that


\be
M^{\rm \scriptscriptstyle inv}_{3} \sim  \C^{abcde}{}_{,2}\,  \partial_a \hat{\Xi}_{b,1} \, J_{cde} \; \; , 
\la{2.16}
\ee

up to $t_0$-exact terms. One can also check explicitly that this expression is not trivial in spinorial cohomology, thereby obtaining a direct proof that this closed super-ten-form really does give rise to the required invariant.

This is the main result of the $D=10$ analysis. We have now shown how all the short $D=4$ invariants can be derived from $D=10$ invariants, and that they are of Chern-Simons type, except for $d^2 \tr^2 F^4$, for which there is a gauge-invariant super-ten-form. Before we move on to discuss the dimensional reduction of these special invariants we shall make some brief comments about other invariants in $D=10$.


\subsection{Other $D=10$ invariants}


The first is the single-trace $d^2 F^4$ invariant. It was shown in \cite{Drummond:2003ex} that, in $D=4$, this is given by the full superspace integral of the Konishi superfield, a result that generalises to all dimensions except $D=10$ where there is no independent Konishi supermultiplet. The supercurrent does admit a non-local decomposition into 128 + 128 together with a scalar superfield \cite{Howe:1982mt}, so that the latter can be projected out. This is the way that this invariant was constructed in \cite{Movshev:2009ba}. However, there is a local alternative, which is to integrate a scalar projection of the Chern-Simons three-form over the whole of superspace. The formula is:

\be
I=\int\,d^{10}x\,d^{16}\th\, (\C^a)^{\a\b} Q_{a\a\b}\ .
\la{2.17}
\ee

Although the integrand is not gauge-invariant, it changes by a total derivative under gauge transformations, and so its integral is an invariant. It clearly has the same dimension as the Konishi superfield and hence will give rise to the same invariant. We shall verify this explicitly when we dimensionally reduce it to nine dimensions in the next section.

There are two further possible invariants of this type. They are given by contracting the $(5,2)$ components of the single- and double-trace Chern-Simons forms with $\C^{5,2}$. These are $d^6 F^4$ invariants; the single-trace example was discussed in \cite{Movshev:2009ba}. In principle one could get invariants of the same type by integrating $\L^4$ over the full superspace, but it is easily seen that there is no scalar in the symmetrised single-trace case owing to the Grassmann nature of $\L$. However, in the double-trace case there is, because the $\L^4$ integrand is just $J^2$. It is likely that the double-trace Chern-Simons seven-form integral is equivalent to this, but we have not explicitly checked this.

In $D<10$, full superspace integrals of $W^4$ give rise to $d^4 F^4$ type invariants. Such invariants cannot be constructed in an obvious fashion in $D=10$ owing to the absence of a scalar field in the SYM multiplet. In the abelian case, it is known that there is such an invariant from superembedding considerations. It is easy enough to reproduce this result using the ectoplasm formalism because we can insert two pairs of contracted derivatives into $H_3 F^4$. This does not generalise to the non-abelian case, at least not in the single-trace case. In fact, it was shown sometime ago \cite{Koerber:2002zb}, using completely different methods, that there should be a single-trace $d^4 F^4$ term in the string effective action and that it goes together with other terms of the same dimension such as $F^6$. Recently, this has been confirmed from a supersymmetric viewpoint: the single-trace $F^4$ term in the action induces higher-order terms through the action of non-linear supersymmetry transformations, and these terms do indeed include $d^4 F^4$ \cite{Howe:2010nu}.\footnote{In \cite{Howe:2010nu} the gauge group is $U(k)$.} So the fact that there is no independent invariant of this type in $D=10$ is perfectly consistent because this term is induced from $F^4$. On the other hand, there should be an independent double-trace $d^4 F^4$ invariant because we can repeat the argument made above for $d^2 F^4$ starting from two pairs of derivative insertions. If this is the case, then it would not necessarily imply that the double-trace $F^4$ term in the action does not induce a $d^4 F^4$ term automatically, although this could be the case if there is only one such term.


\section{Dimensional reduction}


In this section we shall study what happens to the $D=10$ invariants when they are reduced to $D\leq 9$.\footnote{The construction of invariants by dimensional reduction was first discussed in \cite{Deser:1979sx}} We shall see that both of the $F^4$ Chern-Simons invariants retain this property in $D=9$, but below that they can be replaced by gauge-invariant super $D$-forms and that the latter are what one expects from direct computations in these dimensions. We shall underline ten-dimensional vector indices and set $E^{\una}=(E^a,E^r)$, where $r=1\ldots 10-D= n$.  We shall also underline ten-dimensional forms, but continue to write spinor indices as $16$-component. For the SYM potential we have

\be
\unA:= A+ E^r W_r\ ,
\la{3.1}
\ee

where $W_r$ are the scalars in $D$ dimensions. The ten-dimensional field strength is

\be
\unF=(F + G^r W_r) + E^r DW_r -\half E^s E^r [W_r,W_s]\ .
\la{3.2}
\ee

Here $D$ denotes the gauge-covariant exterior derivative in $D$ dimensions, all fields are independent of the compactified coordinates and $G^r=d E^r$ is the dimension-zero torsion in the extra dimensions.
We can easily read off the constraints on the lower-dimensional field-strength tensor from the $D=10$ ones using this equation.

For an arbitrary $p$-form $\uo_p$ we have

\be
\uo_p=\o_p + E^r \o_{p,r} + \ldots \frac{1}{n!} E^{r_n}\ldots E^{r_1}\o_{p-n,r_1\ldots r_n}\ .
\la{3.3}
\ee

If $\uo_p$ is closed, then it is only the last term in the expansion that is closed by itself in the lower dimension.

Now consider the $D=10$ Chern-Simons $F^4$ invariants. They can be derived starting from a closed eleven-form $\unW_{11}=\unH_3 \unP_8$, where ${\unP}_8$ can be single- or double-trace. Moreover, $\unW_{11}= d\unK_{10}$. If we reduce this to nine dimensions, then we find, with $\unW_{11}=W_{11}+ E^9 W_{10}$, and similarly for $K$, that $W_{10}=dK_9$. And we also have

\bea
\unW_{11}&=&W_{11} + E^9 W_{10}\nn\w1
&=&(H_3 + E^9H_2)\unP_8
\la{3.4}
\eea

where $\unH_3=H_3 + E^9 H_2$.  Since $\unP_8=P_8(\unA)$, i.e. is a polynomial determined by $\unA$, and since $\unA=A+ E^9 W$, we know, from the general properties of such polynomials, that $P_8(\unA)=P_8(A) + d( E^9 X_6)$, where $X_6$ is some gauge-invariant six-form in nine dimensions.  We therefore find

\bea
W_{10}&=&H_2 (P_8 + G^9 X_6)-H_3 dX_6\nn\w1
&=&H_2 P_8-d(H_3 X_6)\ ,
\la{3.5}
\eea

since $dH_3=-G^9 H_2$. So the natural $D=9$ Chern-Simons object, $H_2 P_8$, is Weil trivial and is equal to the exterior derivative of $K_9+H_3 X_6$.  We can therefore construct a super-nine-form starting from this point. The lowest component of $K_9$ is $K_{5,4}$, but the lowest component of $H_3 X_6$ actually has bidegree  $(1,8)$, the point being that the natural $F^4$ Chern-Simons construction in $D=9$ leads to a superform whose lowest component is  $L_{1,8}$ as opposed to the $L_{4,5}$ one would have obtained by direct dimensional reduction of the super-ten-form in $D=10$.

We can now repeat this manouevre and go from $D=9$ to $D=8$ starting from the natural $H_2 P_8$  form in $D=9$. Now, however, we find that there is no $H_1$: it is identically zero. Instead, we find that there is a natural closed, gauge-invariant eight-form given by $K_8-H_2 X'_6$, where $K_8$ is derived by dimensional reduction from $K_9+H_3 X_6$ and $X'_6$ is derived from $P_8$ in a similar fashion to  $X_6$, but in one fewer dimension. This means that we now have a gauge-invariant super-eight-form in $D=8$ that gives rise to an $F^4$ invariant by the ectoplasm construction. So the two $F^4$ invariants are of Chern-Simons type in $D=9,10$ but in all lower dimensions they are given by gauge-invariant super-$D$-forms whose lowest component is in $L_{0,D}$. 

The other Chern-Simons type invariant in $D=10$ is the on-shell action which can be obtained starting from the eleven-form $H_7 \tr F^2$. 
In flat superspace $H_7\sim \C_{5,2}$, and so the Chern-Simons nature of the action is maintained for $D\geq 5$. In $D=4$ it reduces to a gauge-invariant expression which is equivalent to the four-$\th$ superaction formula given in \cite{Howe:1981xy}.

We now consider the dimensional reduction of the integral of the Chern-Simons three-form given in \eq{2.17}. We have

\be
Q_3 (\underline{A})=Q_3(A+ E^9 W)=Q_3(A) + E^9 \tr(2 W F(A) + G^9 W^2) +  d (\tr(A E^9 W))\ ,
\la{3.6}
\ee

and, from \eq{3.2} and the $D=10$ constraints, 

\be
F_{\a\b}=i(\C^9)_{\a\b} W\ .
\la{3.7}
\ee

The integrand in \eq{2.17} can be written as

\be
(\C^{\una})^{\a\b} Q_{\una\a\b}(\unA)=(\C^a)^{\a\b} Q_{a\a\b}(A) + (\C^9)^{\a\b} Q_{9\a\b}\ ,
\la{3.8}
\ee

and \eq{3.6} implies that the second term on the right  is proportional to $\tr W^2$ up to a derivative term that integrates to zero. The first term can be evaluated by contracting all four spinor indices in the relation $P_{0,4}=d_1 Q_{0,3} + t_0 Q_{1,2}$  with two factors of $\C^9$. The derivative term will again drop out in the integral, and thus we find, since $P_{0,4}\propto(\C^9)^2 \tr W^2$, that the first term on the right in \eq{3.8} is also proportional to $\tr W^2$. We have therefore shown that

\be
I=\int\,d^{10}x\,d^{16}\th\, (\C^a)^{\a\b} Q_{a\a\b}\rightarrow {\rm const} \int\,d^9x\,d^{16}\th\, \tr W^2
\la{3.9}
\ee

under dimensional reduction to $D=9$. Since $\tr W^2$ is the Konishi superfield in $D=9$ this confirms the claim made below \eq{2.17}.


\section{Implications for ultra-violet divergences}


In the algebraic approach to the renormalisation of MSYM in various dimensions, admissible counterterms are those that can be associated with a closed super-$D$-form that has the same structure as that of the action \cite{Bossard:2009sy}. This result is derived in the component formalism starting from the putative counterterm considered as a spacetime $D$-form and then deducing the existence of a sequence of spacetime $(D-k)$-forms by descent. This descent series is then equivalent to a super-$D$-form of the type we have been considering in the ectoplasm approach.

The $F^4$ invariants, as we have seen, are given by Chern--Simons type  expressions in $D=9,10$ but can be replaced by gauge-invariant super-$D$-forms in $D\leq 8$.  These gauge-invariant super-forms certainly have a distinct structure to that of the action which allows one to prove that counterterms of this type are forbidden within the algebraic approach \cite{Bossard:2009sy}. However, as we saw in section 2.2, the $D=10$ Chern-Simons cocycles for the $F^4$ invariants are already longer than that of the action due to the fact that their lowest components have bi-degrees $(4,6)$ or $(2,8)$ instead of $(5,5)$.  If we reduced these directly, we would expect to retain this structure in lower dimensions. Either way, these invariants would be protected purely by algebraic renormalisation. There is a third argument one could use which is that, because Chern--Simons type operators can only get renormalised by gauge-invariant operators \cite{Breitenlohner:1983pi}, one does not need to consider the possibility that these invariants could be defined as the top component of Chern--Simons type super-$D$-forms of the same structure as that of the action.

Finally, we consider the one-quarter BPS invariant $d^2\tr^2 F^4$. As we have seen, this has a gauge-invariant super-ten-form compatible with the Lagrangian, which can be reduced straightforwardly to any lower dimension. We therefore conclude that the algebraic method does not protect this operator from renormalisation. This is only  relevant  in $D=5,6,7$, since in $D\geq8$ it can only occur in $D=10$ at one loop where the algebraic non-renormalisation theorem is not valid. In $D=7$ this invariant corresponds to a two-loop counterterm which was found to be non-zero in \cite{Marcus:1984ei,Bern:1998ug}. In $D=6$, however, recent computations  \cite{Bern:2010tq} based on the unitarity method have shown that this invariant, which now occurs at three loops, is in fact ultra-violet finite. The algebraic method therefore does not explain this, although it appears that string-based arguments can \cite{Berkovits:2009aw} (see also \cite{Bjornsson:2010wm} for a recent discussion based on a first-quantised pure spinor approach). In \cite{Bossard:2009mn} it was argued that there is a difference between $D=7$ and $D=6$ in that the cocycle for the former could be shown to be equivalent to that of the action while this was not true in $D=6$. One might think that one could simply reduce the $D=7$ expression to $D=6$, but this would not automatically lead to a  fully $D=6$ R-symmetric super-form. However, a closer analysis of the cohomology reveals that there is not a problem with this. The cohomological analysis in $D=6$ and $D=5$ is somewhat involved, but the construction of the invariant in $D=10$ simplifies the analysis considerably as it is now obvious that the one-quarter BPS invariant is not protected by the algebraic method in any dimension.

The finiteness of $D=6$ MSYM at three loops in the double-trace sector therefore remains a puzzle from a purely field-theoretic point of view. It seems difficult to envisage a mechanism that can allow a counterterm in one dimension but not in another based on symmetry principles. There is one difference between $D=6$ and $D=7$, however. In $D=6$ this invariant is strictly one-quarter BPS, while in $D=7$, as we mentioned in the introduction, this notion is not compatible with Lorentz symmetry.  If one takes the product of two supercurrents, for $D\leq 6$, the lowest scalar components define four different multiplets: one-half BPS, one-quarter BPS, and non-BPS, long and short types. This last multiplet satisfies a higher-order differential constraint (in terms of $D$s). For example, in $D=4$, the square of the supercurrent in the real $20$-dimensional representation of $SU(4)$ satisfies a second-order constraint which leads to its being a protected operator even though it is not BPS \cite{Heslop:2001dr}.\footnote{This operator was found to be non-renormalised by direct calculation in the context of AdS/CFT in refs \cite{Arutyunov:2000ku,Eden:2000bk}.} In $D=7$, on the other hand, the square of the supercurrent splits into three scalar superfields, of which one is one-half BPS, one is long, and the third obeys a higher-order constraint. It is this third one that contains the $d^2\tr^2 F^4$ invariant. It may be the case that the sum of the one-quarter  BPS and short multiplets is no longer reducible in $D=7$, or perhaps that one can separate them at higher dimension. Either way, it does seem that there is a real difference but it is not clear that the algebraic method can detect this in the analysis of divergences. It might be that there is some, as yet unknown, off-shell formalism that is subtly different in the different dimension, but this is purely conjectural.


\section{Summary and conclusions}


In this article we have studied the invariants of maximally supersymmetric Yang-Mills theory with gauge group $SU(k)$, in $D=4,\ldots 10$ using superspace techniques. We have focused on the short invariants because the long invariants, corresponding as they do to full superspace integrals, are much simpler to deal with. As we have seen, for $D\leq 6$, all of the short invariants are of BPS type except for the on-shell action itself, and this is also expressible as a gauge-invariant superaction in $D=4$. There does not seem to be any good reason to expect this situation to change in $1\leq D\leq 3$, although we have not investigated this fully. For $D>6$, the multiplet structure starts to change, and in particular, in $D=9,10$ we have seen that the $F^4$ invariants are of Chern-Simons type. We have also shown how one can construct further invariants by inserting spacetime derivatives and have made use of this trick to express the $d^2\tr^2 F^4$ invariant in terms of a gauge-invariant closed super-ten-form via the ectoplasm formalism. In \cite{Movshev:2009ba} a classification of single-trace invariants, mainly in $D=10$, was given. The authors of this paper used completely different cohomological techniques to the ones we have used here, and in particular, did not make the connection between invariants and short multiplets so explicit. 

The results of our investigations are summarised in the following table:

\vskip .5cm

\begin{table}[ht]
\newcommand\TOP{\rule{0pt}{4ex}}
\begin{center}
\begin{tabular}{|c||c|c|c|c|c|}
\hline
\TOP & $\mathrm{tr}\,F^2$ & $\mathrm{tr}\,F^4$,\  $\mathrm{tr}^2\,F^4$ & $d^2\mathrm{tr}^2\,F^4$ &$d^2\mathrm{tr}\,F^4$  \\[5pt]
\hline\hline
\TOP $D=10$ & $\int\,L^{CS}_{D,0}$  & $\int\,L^{CS}_{D,0}$ & $\int\,L^{\rm \scriptscriptstyle inv}_{D,0}$ & $\int d^{10}x\,d^{16}\theta\,Q$ \\[5pt]
\hline
\TOP $D=9$  &  '' & ''& '' & $\int d^Dx\,d^{16}\theta\,K$ \\[5pt]
\hline
\TOP $D=7,8$ &'' & $\frac{1}{2}$ BPS & '' & '' \\[5pt]
\hline
\TOP $D=5,6$ & ''& ''&$\frac{1}{4}$ BPS&''\\[5pt]
\hline
\TOP $D=4$&$\half$ BPS &''&''&''\\[5pt]
\hline
\end{tabular}

\vskip .5cm

\caption{Some MSYM invariants and their construction in various dimensions. The short invariants are, besides the action, the single and double trace $F^4$ invariants as well as the double trace $d^2F^4$ invariant. $Q$ denotes the projection of the CS three-form and $K$ is the Konishi superfield. The one-half BPS integral for the action in $D=4$ is special (ultra-short) in that it is an integral over four $\th$s rather than $8$. For $1\leq D\leq3$ the situation is expected to be the same as in $D=4$.}
\end{center}
\end{table}

Combining these results with algebraic renormalisation techniques we were able to confirm that the single- and double-trace $F^4$ invariants cannot mix with the action, owing to the fact that they have different cocycle structures. On the other hand, the existence of a gauge-invariant closed super-ten-form for the $d^2\tr^2 F^4$ in $D=10$ dimensions implies that the cocycle structure associated with this invariant will be the same as that of the action in all lower-dimensional spacetimes and hence that one is not able to conclude that it is protected by this argument even though it is known that it is finite in $D=6$. There is a difference for this invariant in $D=7$ in that it is only truly one-quarter BPS in $D\leq 6$ and below, but it is not clear from the algebraic point of view why this distinction should affect its ultra-violet properties.

\end{document}